\documentstyle[sprocl]{article}

\input{epsf}

\def\Journal#1#2#3#4{{#1} {\bf #2}, #3 (#4)}

\def\NPB{{\em Nucl. Phys.} B}
\def\PLB{{\em Phys. Lett.}  B}
\def\PRL{\em Phys. Rev. Lett.}
\def\PRD{{\em Phys. Rev.} D}

\def\be{\begin{equation}}
\def\ee{\end{equation}}
\def\bea{\begin{eqnarray}}
\def\eea{\end{eqnarray}}

\begin{document}

\title{ END POINT OF THE ELECTROWEAK PHASE TRANSITION \footnote{Presented by 
Z. Fodor at SEWM'98, Copenhagen}}

\author{F. CSIKOR, Z. FODOR}

\address{Institute for Theoretical Physics, E\"otv\"os University,\\
P\'azm\'any P\'eter s\'et\'any 1A, H-1117 Budapest, Hungary
\\E-mail:fodor@thwgs01.cern.ch } 

\author{J. HEITGER}

\address{DESY Zeuthen, \\
Platanenallee 6, D-15738 Zeuthen, Germany }

\author{Y. AOKI, A. UKAWA}

\address{Center for Computational Physics, University of Tsukuba,
  Tsukuba, Ibaraki 305-8577, Japan\\
  Institute of Physics, University of Tsukuba,
    Tsukuba, Ibaraki 305-8571, Japan }


\maketitle\abstracts{ 
We study the hot electroweak phase transition (EWPT) by 4-dimensional lattice
simulations on lattices with symmetric and asymmetric lattice spacings 
and give the phase diagram. A continuum
extrapolation is done. We find first order phase transition 
for Higgs-boson masses $m_H<66.5 \pm 1.4$ GeV. 
Above this end point a rapid cross-over occurs. 
Our result agrees with that of the dimensional
reduction approach. It also indicates that the fermionic
sector of the Standard Model (SM) may be included perturbatively.
We get for the SM end point  $72.4 \pm 1.7$ GeV.
Thus, the LEP Higgs-boson mass lower bound excludes
any EWPT in the SM.
}

\section{Introduction}
\noindent
The observed baryon asymmetry is finally determined at the
EWPT \cite{KuRS}.
To understand this asymmetry a quantitative
description of EWPT is needed. Since
the perturbative approach breaks down for large
Higgs-boson masses \cite{pert} (e.g. $m_H>70$ GeV) 
lattice MC simulations are necessary. 
(Another, partly analytic approach has been presented in \cite{Meyer}.)

Previous works show that the strength of the 
EWPT gets weaker as the Higgs-boson mass 
increases. The line of the first order phase
transitions, separating the symmetric and broken phases
on the $m_H-T_c$ plane, has an end point, $m_{H,c}$. 
3D results show that for $m_H>95$ GeV
no EWPT exists, \cite{3d95}  moreover
 $m_{H,c} \approx 67$ GeV \cite{3d80,Gurtl,Rum}. 
In 4D \cite{Aoki1} $m_{H,c} \approx 80$GeV was obtained. 
Also in  4D at $m_H \approx 80$ GeV the EWPT turned out
to be extremely weak, even consistent with the no phase
transition scenario on the 1.5-$\sigma$ level \cite{4d80}.
However, a discrepancy appeared: the 4D estimate for the end point
was higher than the 3D estimates.
This discrepancy has been resolved in \cite{Aoki3,CS3},  
to be reviewed here. 
\begin{figure}[t]
\centerline{\epsfxsize=0.50 \linewidth \epsfbox{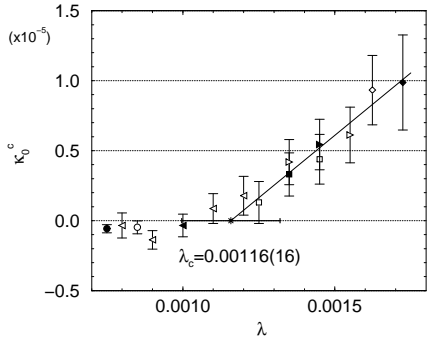}}
\caption{\label{fig0}
{ Imaginary part of first Lee-Yang zero 
as a function of $\lambda$  
from simulations on symmetric lattices with $L_t=2$ \protect\cite{Aoki3}. 
Filled symbols are obtained without $\lambda$-reweighting,
while open symbols with
$\lambda$-reweighting from the filled symbol with same shape.
}}
\end{figure}

\section{End point analysis}
\noindent
We  study the 4D SU(2)-Higgs model
on both symmetric \cite{Aoki3}  and asymmetric \cite{CS3} 
lattices, i.e. lattices
with equal or different spacings in temporal ($a_t$) and spatial
($a_s$) directions. In the asymmetric case 
equal lattice spacings are used in the 3
spatial directions ($a_i=a_s,\ i=1,2,3$). 
The asymmetry of the
lattice spacings is given by the asymmetry factor $\xi=a_s/a_t$.
The different lattice spacings can be ensured by
different coupling strengths in the action for time-like and space-like
directions. The action reads in standard notation \cite{4d-rev}
\begin{eqnarray}\label{lattice_action}
&&S[U,\varphi]= 
 \beta_s \sum_{sp}
\left( 1 - {1 \over 2} {\rm Tr\,} U_{pl} \right)
+\beta_t \sum_{tp}
\left( 1 - {1 \over 2} {\rm Tr\,} U_{pl} \right)
+ \sum_x \left\{ {1 \over 2}{\rm Tr\,}(\varphi_x^+\varphi_x) \right.
\nonumber \\ &&  +
\lambda \left[ {1 \over 2}{\rm Tr}(\varphi_x^+\varphi_x) - 1 \right]^2
- \left. \kappa_s\sum_{\mu=1}^3
{\rm Tr}(\varphi^+_{x+\hat{\mu}}U_{x,\mu}\,\varphi_x)
-\kappa_t {\rm Tr\,}(\varphi^+_{x+\hat{4}}U_{x,4}\,\varphi_x)\right\}.
\end{eqnarray}
We introduce  
$\kappa^2=\kappa_s\kappa_t$ and
$\beta^2=\beta_s\beta_t$. The anisotropies
$\gamma_\beta^2=\beta_t/\beta_s$ and $\gamma_\kappa^2=\kappa_t/\kappa_s$
are functions of $\xi$. These functions have been
determined perturbatively and non-perturbatively
\cite{T0nonpert} demanding the restoration of the rotational
symmetry in different channels. We use 
$\xi=4.052$, which corresponds to $\gamma_\kappa=4$ and $\gamma_\beta=3.919$.
Details of the simulation techniques can be found in \cite{4d-rev}.
\begin{figure}[t]
\centerline{\epsfxsize=0.50 \linewidth \epsfbox{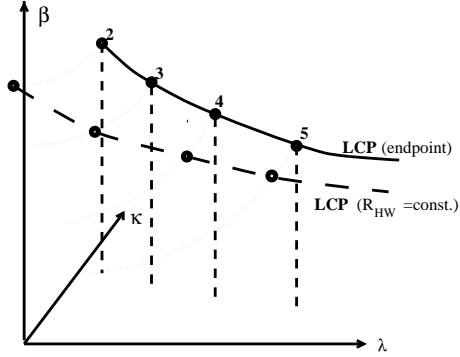} }
\caption{\label{fig1}
{ Schematic phase diagram. The solid line represents the
LCP defined by the end point condition. The numbers on the line correspond to
the temporal extension  (the dashed
lines show  their projection to the $\kappa$ - $\lambda$ plane). The dotted
lines running into these numbered points correspond to first order phase
transitions for $g_R^2$ = const. but different $R_{HW}$-s. A LCP defined
by a constant $R_{HW}$ value is shown by the long dashed line.
}}
\end{figure}

The determination of the end point of the finite temperature
EWPT is done by the use of the Lee-Yang zeros \cite{LY} of the
partition function ${\cal Z}$.
Near the first order phase transition point the partition function reads
${\cal Z}={\cal Z}_s + {\cal Z}_b \propto \exp (-V f_s ) + \exp ( -V f_b ) \, ,
$
where the indices s(b) refer to the symmetric (broken) phase and $f$ stands
for the free-energy densities. We also have
$f_b = f_s + \alpha (\kappa - \kappa _c ) \,$ ,
since the free-energy density is continuous. It follows that
${\cal Z} \propto  \exp [ -V ( f_s +f_b )/2 ] 
\cosh [ -V \alpha (\kappa -\kappa_c )] \,
$
which shows that for complex $\kappa$ ${\cal Z}$ vanishes at
${\rm Im} (\kappa )=2 \pi \cdot (n-1/2) / (V\alpha )$
for integer $n$.  In case a first order phase transition is present,
these Lee-Yang
zeros move to the real axis as the volume goes to infinity. In case a
phase transition is absent the Lee-Yang
zeros stay away from the real $\kappa $ axis. Thus the way the Lee-Yang
zeros move in this limit is a good indicator for the presence or
absence of a first order phase transition \cite{LY}.
Denoting
$\kappa_0$ the lowest zero of ${\cal Z}$, i.e. the  position of the zero
closest to the real axis, one expects in the vicinity of
the end point the scaling law
${\rm Im}(\kappa_0)=C(L_t,\lambda)V^{-\nu}+\kappa_0^c(L_t,\lambda)$.
In order to pin down the end point we are looking
for a $\lambda$ value for which $\kappa_0^c$ vanishes.
In practice we analytically continue ${\cal Z}$ to complex
values of $\kappa $ by reweighting \cite{FS} the available data.
Also small changes  in
$\lambda$ have been taken into account by reweighting.
The dependence of $\kappa_0^c$ on  $\lambda$ from our symmetric simulations
\cite{Aoki3} is shown in Fig.~1.
To determine the critical value of $\lambda$
i.e. the largest value, where $\kappa_0^c=0$, we
have performed  fits linear in $\lambda$ to the non-negative 
$\kappa_0^c$ values.
\begin{figure}[t]
\centerline{\epsfxsize=0.50 \linewidth \epsfbox{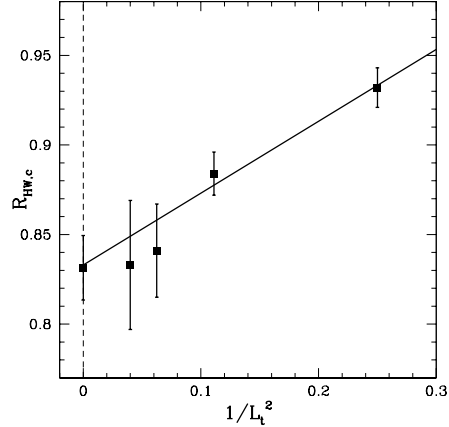}}
\caption{\label{fig3}
{ Dependence of $R_{HW,c}$, i.e. $R_{HW}$ corresponding to the end point
of first order phase transitions, on $1/L_t^2$ and extrapolation to the
continuum limit.
}}
\end{figure}

In the isotropic lattice simulation \cite{Aoki3}, we have used $L_t =2$. 
The Lee-Yang analysis gave $\lambda_c =0.00116(16)$ for the end point. 
Performing $T=0$ simulations with the same parameters this can be converted 
to a Higgs boson mass. The value $m_{H,c}=73.3 \pm 6.4$GeV has been obtained, 
which is compatible with our estimate based on a study of 
Binder cumulants \cite{Aoki3} 
and the previous 4D studies  \cite{Aoki1,4d80}.

In the anisotropic lattice simulation case \cite{CS3} 
we also performed a continuum extrapolation,
moving along the lines of constant physics (LCP). A schematic 
illustration is shown in Fig. 2.

The technical implementation of the LCP idea has been done as follows.
By fixing $\beta=8.0$ in the simulations, we have observed that  $g_R $ is
essentially constant within our errors.  For the small differences in $g_R $
we have performed perturbative corrections. We have carried out $T \neq 0$
simulations  on $L_t= 2,3,4,5$ lattices (for the finite temperature case
one uses $L_t \ll L_x ,L_y ,L_z )$, and tuned $\kappa $ to the transition
point.  This condition fixes the lattice spacings: $a_t =a_s / \xi \, = \,
1/(T_c L_t ) $ in terms of the transition temperature $T_c $ in physical units.
The third parameter $\lambda $, finally specifying the physical Higgs mass
in lattice units, has been chosen   (using the Lee-Yang analysis) 
so that the transition
corresponds to the end point of the first order phase transition subspace.

Having determined the end point
$\lambda_c (L_t)$ for each $L_t$ we calculated the $T=0$
quantities ($R_{HW},g_R^2$) on $V=(32L_t)\cdot (8L_t)\cdot (6L_t)^3$ lattices.
Having established the correspondence
between $\lambda_c (L_t)$ and $R_{HW,c}$, the $L_t$ dependence of the
critical $R_{HW,c}$ is easily obtained. Fig. 3 shows the dependence of
the end point $R_{HW}$ values on $1/L_t^2 $. A linear extrapolation
in $1/L_t^2 $ yields the continuum limit value 
$R_{HW,c} = 0.83 \pm 0.02$, which corresponds  to $m_{H,c} =66.5 \pm 1.4$ GeV.
Note that $m_{H,c} $ decreases for increasing $L_t$. This observation 
resolves the discrepancy of 3D \cite{3d80,Gurtl,Rum} and previous $L_t=2$ 4D 
\cite{Aoki1,4d80} results.

Comparing our result to those of the 3D analyses \cite{3d80,Gurtl,Rum} one
observes complete
agreement. Since the error bars on the end point determinations are on the
few percent level, the uncertainty of the dimensional reduction procedure
is also in this range. This indicates that the analogous perturbative
inclusion  of the fermionic sector results also in few percent error on
$m_{H,c}$.

\begin{figure}[t]
\centerline{\epsfxsize=0.50 \linewidth \epsfbox{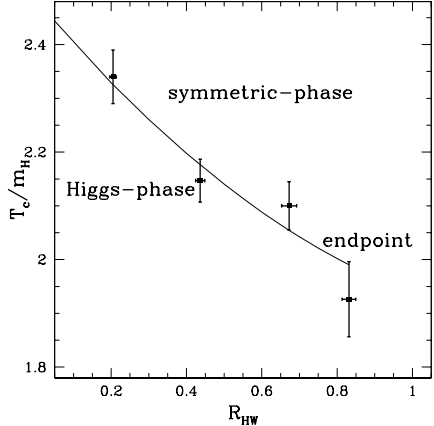}}
\caption{\label{fig3v}
{  Phase diagram of the SU(2)-Higgs model in the ($T_c /m_H - R_{HW} $)
plane. The continuous line -- representing the phase-boundary -- is a quadratic
fit to the data points.  }}
\end{figure}

Based on our published data \cite{4d-rev,T0nonpert} and the results of 
\cite{CS3} we draw the precise phase diagram in the ($T_c /m_H - R_{HW} $) 
plane of the SU(2)-Higgs model. This is shown in
Fig. 4. The continuous line -- representing the phase-boundary -- is
a quadratic fit to the data points.

Finally, we determined the end point value in the full SM.
We use perturbation
theory \cite{KLRS96} to transform the SU(2)-Higgs model end point value to
the full SM. We obtain $72.4 \pm 1.7$ GeV. The dominant error
comes from the uncertainty on the position of the end point.

\section{Conclusions} 
\noindent
In conclusion, we have determined the end point of hot EWPT
with the technique of Lee-Yang zeros from simulations
in 4D SU(2)-Higgs model. The phase diagram has been also
presented. The transition is first order for Higgs masses less than
$66.5 \pm 1.4$ GeV, while for larger Higgs masses only a rapid cross-over
is expected. Our results show   
that integrating out the heavy modes  perturbatively is sufficiently precise.
Thus the above value can  be perturbatively  transformed  to
the full SM, yielding $72.4 \pm 1.7$ GeV for the end point
Higgs mass. 

The experimental lower limit of the SM Higgs-boson mass
is $89.8$ GeV \cite{LEP}. Taking into account all errors, 
our end point value excludes the
possibility of any EWPT in the SM.
Thus our work emphasizes the need of EWPT analyses  
based on extensions of the SM \cite{MSSM}.

\section*{Acknowledgements}
Work supported by Min. of Education, Japan (No. 10640246),
Hung. Sci. Foundation (No. OTKA-T016240/T022929) and
FKP-0128/1997.

\end{document}